# Gaming the Attention Economy[1]

Daniel Estrada, University of Illinois, Urbana-Champaign
djestrada@gmail.com

Jonathan Lawhead, Columbia University
reality.apologist@gmail.com

**Abstract:**

The future of human computation (HC) benefits from examining tasks that agents already perform and designing environments to give those tasks computational significance. We call this *natural human computation* (NHC). We consider the possible future of NHC through the lens of Swarm!, an application under development for Google Glass. Swarm! motivates users to compute the solutions to a class of economic optimization problems by engaging the attention dynamics of crowds. We argue that anticipating and managing economies of attention provides one of the most tantalizing future applications for NHC.

## 1. Natural Human Computation

Human computation (HC) involves the creation of mixed organic-digital systems to solve difficult problems by outsourcing certain computational tasks to the human brain. However, we can distinguish between HC approaches that require a user to engage with a specific (and arbitrary) program or system, and HC approaches that simply leverage a user's normal activity to compute the solutions to complex problems. We call this latter approach *natural human computation* (NHC). An instance of HC is *natural* when the behavior necessary for carrying out the proposed computation is already manifest in the system.

Eusocial insect colonies are models of natural computation (see Gordon, 2010; Moses et al. and Pavlic and Pratt, this volume). The information processing potential of ant colonies emerges from the small-scale, everyday interactions of individual ants: everything individual ants do is computationally significant, both for the management of their own lives and for the colony's success. This alignment between individual and colony-level goals means that ant colonies need not direct the behavior of individual ants through any sort of top-down social engineering. The queen issues no royal decrees; insofar as she has any special control over the success of the colony, that control is a product of her influence on individual colony members with whom she comes into contact. The sophisticated information processing capabilities of the colony as a whole are a product of each ant obeying relatively simple local interaction rules--those local interaction rules, however, allow an aggregate of ants to influence each others' behavior in such a way that together, they are capable of far more complicated computing tasks than individual colony members would be on their own.  Crucially, the computational power of the colony *just is* the concerted action of individual ants responding to the behavior of other ants: any change in the colony's behavior will both be a result of and have an impact on the behavior of colony members. In this sense, natural ant behavior is both *stable* and *natural:* the computing activity of the colony can't disrupt the behavior of colony members out of their standard behavior routines, since those standard behavior routines *just are* the computing activity of the colony. The stability of this behavior can in turn support a number of additional ecological functions. The regular harvesting of individual bees not only supports the activity of the hive, but also solves the pollination problem for flowers in what we might call "natural bee computing"[2] which piggybacks on the behavior. NHC approaches take these natural models of computation as the paradigm case, and seek to implement similar patterns in human communities.

---

[1] Our sincere thanks to Pietro Michelucci for his prompt, helpful, and encouraging comments on drafts of this paper. His patience and assistance in its production has been invaluable.

We have sketched a definition for NHC in terms of *stable* and *disruptive* computation, and turn now to discuss these concepts directly.  Disruptive computation requires a *change* in an agent's behavior in order to make their performance computationally significant. Human computation is increasingly *stable* as its impact on agent behavior is reduced. Describing an instance of human computation as "natural" is not itself a claim that the *human activity* is stable or disruptive, since NHC techniques can be used to to extract computationally significant data in both stable and disruptive contexts. Rather, describing an instance of HC as natural makes the more limited claim that the computation in question was not *itself* a source of disruption. We introduce the vocabulary of stability and disruption to clearly articulate this aspect of NHCs.

It may be instructive to compare NHC and gamification (Deterding et al., 2011; McGonigal, 2011) as strategies for human computing.  Gamification makes an HC task more palatable to users, but often alters user behavior in order to engage with the computational system. In contrast, NHC systems transparently leverage existing behaviors for computation. For instance, reCAPTCHA (von Ahn et al., 2008) repurposes an existing task (solving text-recognition puzzles to gain access to a website) to solve a new problem (digitizing books for online use).  This pushes HC to the background; rather than explicitly asking users to participate in the solution of word recognition problems, it piggybacks on existing behavior.  Gamification is not always disruptive in the sense used here; in some cases described below gamification techniques can serve to *stabilize* (rather than *disrupt*) the dynamics of systems to which they are applied.  This suggests that we need a more robust vocabulary to map the conceptual territory.

Michelucci (this volume) distinguishes between "emergent human computation" and "engineered human computation."  Emergent HC systems analyze uncoordinated behavior from populations to do interesting computational work, while engineered HC systems might be highly designed and coordinated for specific computing needs. We see natural human computation as a concept that is complementary to but distinct from Michelucci's distinction.  The defining characteristic of NHC is the potential for extracting additional computational work from human activity without creating additional disturbances in that behavior. This definition makes no assumptions about the degree to which these behaviors have been designed or coordinated for particular computing functions. In fact, we assume that natural human behavior involves organizational dynamics that cut across Michelucci's distinction. NHC systems like Swarm!, described in Section 2 below, can be understood as a method for discerning natural organizational patterns as a potentially fruitful source of human computation.

---

[2] Of course, bees and flowers achieved this stable dynamic through millions of years of mutualistic interaction; as we discuss in section 4, we expect any HC technique to require some period of adaptation and development.

We're thinking about NHC in terms of the impact a computing task has on the behavior of its computers; NHC tasks introduce minimal disruptions to existing behavior. In contrast, Michelucci's distinction isn't concerned with the impact HC has on its agents. Rather, it is concerned with the performance of the computing task in question. Emergent cases of computing are where the goal is best accomplished by passively analyzing agents for specific computational results, more or less independent of other aspects of their behavior. Engineered systems require increasingly coordinated activity to achieve computational results. For these reasons, we consider Michelucci's distinction to be a system-level or "top-down" perspective on computing tasks, while the stable/disruptive distinction is an agent-level or "bottom-up" perspective on the same tasks. Or to cast the issue in techno-phenomenological terms: Michelucci is taking a designer's perspective on human computing, where purposes (functions, tasks, goals, ends) are *imposed* on a computing population; on the other hand, we're interested in the user's perspective, where the generation and pursuit of purposes is a constitutive aspect of one's ongoing committed engagement with the world.

It is worth reiterating that the sense of "natural" being articulated cuts across the categories represented in Table 1 below. We can think of these categories as defining the axes of a continuous space of possible computing systems. Claiming that a given system is emergent and disruptive (for instance) is to locate within this space. However, claiming that a given instance of human computation is *natural* is to point out a very different sort of fact about the system. In the context of human computation, *naturalness* is something like an indexical, describing words with use-relative content like "here" or "now." Rather than giving an absolute location in the space defined by the distinctions discussed above, calling an instance of HC "natural" is to assert a fact about the HC system *relative* to the current state of the computational substrate. A NHC might be engineered, emergent, disruptive, or stable to some greater or lesser degree; the ascription of naturalness depends only on a comparison between the system's state *now* and the state that would be necessary for performing the desired computation. The distinctions between emergent, engineered, stable, and disruptive HC systems can be more clearly illustrated if we consider a few representative examples. An absolute attribution of naturalness in any of these cases is not possible, as "naturalness" is an index to a user-relative state. As such, the following examples contain no direct appeal to "naturalness", since the degree of naturalness for some HC process may vary between individual users with distinct behavioral profiles. Using Yelp in deciding on some service, or using ZR to motivate your run, will integrate naturally into the usage patterns of some users and may be more disruptive in the lives of others.

Consider the following cases:

|  | Stable | Disruptive |
|---|---|---|
| **Emergent** | American Idol predictions | Yelp |
| **Engineered** | Zombies Run | FoldIT |

**Table 1**

**Emergent/Stable:** HC systems are emergent when they exploit uncoordinated behavior in a population, and they are stable when that computing goal is met without further disruption. reCaptcha has already been mentioned as an example of HC that falls in this quadrant. A more illustrative example can be found in Ciulla et al. (2012), which describes modeling approaches to the Twitter datastream that successfully anticipate the results of a recent American Idol voting contest. In this study, users Tweeted their thoughts on the contest of their own accord[3], without coordination and independently of their potential use in predictive speculation, and so meets the definition of emergent. Solving the prediction task required no additional input from the users beyond this existing social behavior, and so also meets the definition of stable.

**Engineered/Stable:** Engineered computing tasks are highly coordinated and designed for specific computing purposes. These designs can be stable in our sense when the computation fits existing patterns of behavior rather than creating new ones. BOINC's successful @HOME distributed computing projects (Anderson 2004) are familiar examples of stable computing strategies, using spare processor cycles for useful computational work without occupying an additional computational footprint. For a more explicitly gamified example, consider the 2012 exercise motivation app called "Zombies Run"[4]. Zombies Run (ZR) is designed to work in tandem with a player's existing exercise routine, casting her as a "runner" employed by a post-apocalyptic settlement surrounded by the undead. The game's story is revealed through audio tracks rewarding player for gathering supplies, distracting zombies, and maneuvering through the dangerous post-apocalyptic wasteland, all accomplished by monitoring a few simple features of the user's run. The app motivates runners to continue a routine they've already developed, using tools already appropriated in that behavior; the app isn't designed to help users to start running, it is designed to help them *keep* running. This is a defining feature of engineered/stable systems: while they are the product of deliberate design, the design's primary effect is to reinforce (rather than alter) existing patterns of behavior. While ZR players aren't (necessarily) performing any particularly interesting computing function, the app provides a clear example of how a highly designed, immersive app can nevertheless be stably introduced into a user's activity.

---

[3] We ignore for the sake of the example any potential feedback from advertising or other systems that reinforce tweeting behavior surrounding the American Idol event.
[4] From the UK-based Six to Start. https://www.zombiesrungame.com/

**Emergent/Disruptive:** A computational state is *disruptive* when implementation would involve a significant reorientation of the behavior and/or goals of the agents under consideration. This can occur in emergent computing contexts where individuals are acting independently and arbitrarily. Yelp.com is a popular web-based service that compiles crowd-sourced reviews of local businesses and services. These reviews are used to compute a rating of a given service based on search criteria. And indeed, solving this computing problem itself changes the activity of the population: Luca (2011) finds that the a one-star rating increase amounts to a 5-9 percent increase in revenue. In other words, the self-directed, emergent activity of Yelp reviewers is disruptive to the behavior of the dining community, effectively redirecting a portion of them to services with higher ratings. It may be supposed that Yelp's disruptive status is a consequence of feedback from the HC system being used to guide the decisions of future diners. However, Zombies Run provides an example where feedback on HC behaviors can reinforce those behaviors with little disruption. This suggests that Yelp's economic impact involves more than providing feedback on the HC task; it reflects something about the specific computations performed by the system. We will return to this point in section three.

**Engineered/Disruptive:** FoldIT is a puzzle-solving game in which the puzzles solved by players are isomorphic to protein folding problems (Khatib et al. 2011). FoldIT is a paradigm case of gamification: it makes a HC task more palatable to the users, but significantly disrupts their behavior in the process by demanding their focus on the game. FoldIT is engineered in the sense that the task has been deliberately designed to provide computationally significant results, and disruptive in the sense that the task is a departure from the behavior in which players otherwise engage.

The above examples are offered in the hopes of making clear a complex conceptual landscape that serves as the backdrop for the discussion of natural human computing. A full discussion of the dynamics of purposive human behavior is beyond the scope of this paper, but we understand our contributions here as a step in that direction. Despite the perspectival dimensions of "naturalness," we can talk sensibly about designing natural human computing systems that leverage existing HC work in minimally disruptive ways. We turn now to describe a NHC system that demonstrates these features.

## 2.0 Introducing Swarm!

Swarm!, an application under development for Google Glass[5], is an implementation of NHC methods for solving a class of economic optimization problems. Swarm! uses the GPS coordinates of players to construct a location-based real time strategy game that users can "play" simply by going about their everyday routines. Individual cognitive systems have limited resources for processing data and must allocate their attention (including their movement through space and time) judiciously under these constraints. Therefore, we can interpret the data gathered by Swarm! as a NHC solution to the task of attention management: Swarm! generates a visualization of aggregate human activity as players negotiate their environments and engage objects in their world.

---

[5] Glass is a wearable computer designed and manufactured by Google. The Glass headset features a camera, microphone with voice commands, optical display, and a touch-sensitive interface. It duplicates some limited functions of a modern smartphone, but with a hands-free design. Fig. 1 depicts a user wearing a Google Glass unit.

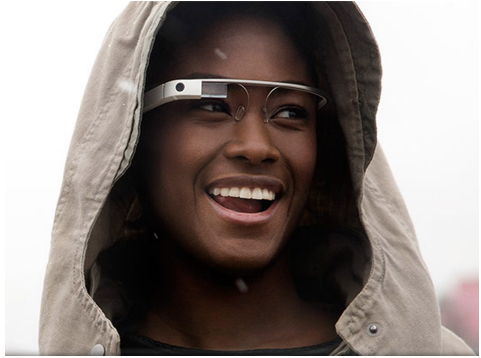

**Fig. 1**

The Swarm! engine is designed as a basic NHC application: it's a game that's played just by going about your normal routine, frictionlessly integrating game mechanics into a player's everyday life.  Swarm![6] simulates membership in a functioning ant colony, with players assuming the role of distinct castes within one colony or another.  Players are responsible for managing their own resources and contributing to the resource management of the colony. Swarm! data is visualized as colorful trails on a map card presented on request to the Glass display in order to engage the resulting behavior. These trails are designed so they can not be used to locate or track any individual uniquely. Instead, we're interested in the broader patterns of behavior: where do players spend their time?  When is a certain park most likely to be visited? When and where do players from two different neighborhoods cross paths most frequently?

## 2.1  Swarm! mechanics

Ant behavior is coordinated through purely local interactions between individuals and a shared environment without any central direction (Dorigo, 2000).  Individual ants exchange information primarily through direct physical contact and the creation of pheromone trails. Pheromone trails, which can be used to indicate the location of resources, warn of danger, or request help with a tricky job, are temporary (but persistent) environmental modifications laid down by individual that help ants coordinate with each other and organize over time to manage colony needs.

Swarm! adopts the pheromone trail as its central mechanic.  By moving around in physical space, players lay down "trails" that are visible through the in-game interface as colorful lines on a map.  These trails encode context-specific information about the history and status of user interactions around a location.  Just like real-world ants, Swarm! trails are reinforced by repeated interaction with a region of space, so the saturation of trails in a particular location represents the degree of activity in that location.  Trails also encode some information about in-game identity, but the focus of Swarm! is on impersonal aggregate data and not unique player identification.  Since trails are semi-persistent and fade slowly with time, the specific time that a player passed a location cannot be deduced by looking at the map.  Players also have the option to define "privacy zones" around their homes and other sensitive areas where Swarm! data collection is prohibited.

---

[6] Complete game bible can be found at http://www.CorporationChaos.com

Swarm! gameplay is styled after many popular resource collection games, with central goals revolving around finding enough food to stay alive, disposing of trash ("midden"), and defending the colony from incursions by rivals.  However, Swarm!'s central innovation is its emphasis on self-organized dynamic game maps and frictionless player interaction. Player interactions result primarily from trail crossings: when one player crosses the trail laid down by another player, an appropriate context-dependent event is triggered.  Note that this activity does not require players to be present simultaneously at one location. Trails laid down by users decay gradually over time, and require reinforcement to sustain. Thus, crossing the trail of a rival ant means that ant (or possibly several ants from the same colony) have reinforced this trail within the decay period. In other words, all player activity is rendered on the map as "active" and will trigger engagements and events specific to those interactions.

Players also have caste-specific abilities to augment the structure of the game map. These abilities are triggered by more in-depth interaction with a location--for instance, spending an extended amount of time in the same public place, or taking some number of pictures of an important game location.  Each caste has a unique set of strengths, weaknesses, and abilities that affect the range of in-game options available to the player.  These augmentations can provide powerful bonuses to members of a player's colony, hinder the activities of rivals, or alter the availability of resources in the area.  Strategic deployment of these abilities is one of the most tactically deep and immersive aspects of Swarm! gameplay.

For illustration, consider the following in-game scenario (Fig 2). Suppose a player (call her Eve) consistently moves through territory that is controlled by an enemy colony--that is, she crosses a region that is densely saturated with the trails of hostile players. Moving through this region has a significant negative impact on Eve's resource collection rate, and unbeknownst to Eve (who doesn't like to be bothered by game updates) this penalty has been adversely affecting her contributions to her colony for weeks, keeping her at a relatively low level than where she might be otherwise. However, suppose that one day Eve decides to actively play Swarm!. Upon downloading the latest game map she observes the impact this region has had on her collection rate. Swarm!'s game mechanics reward this attention to detail, and allow Eve to do something about it. When Eve photographs the locations that are controlled by a rival colony, she creates an in-game tag that calls attention to her predicament and provides caste-specific in-game effects that potentially offset the impact of the rival colony's trail. In other words, her action (taking a picture) has produced an in-game structure that warps the map and partially ameliorates the penalty that she would otherwise suffer. This in-game structure might attract other active players to the territory to build more structures that further magnify these changes. In this way, close attention to (and interaction with) the game map is rewarded, while casual players are still able to contribute meaningfully to the overall game dynamic.

This reveals an important aspect of Swarm! related to the distinctions drawn in Section **1**. Although the game is designed to passively harvest aggregate user behavior, it also incentivizes the curation of that data allowing for active user engagement. Thus, some users may experience Swarm! as unobtrusive and stable, with computation occurring largely in the background, while others may enjoy significant disruptions as they actively play the game. Moreover, the two might interact with each other through in-game mechanics around shared spaces without either player being aware of the other's presence. When Eve tags a highly trafficked area of the map with her picture, she is highlighting an attractor[7] in *both* the physical space and the game space. Those attractors emerge naturally in the behavior of some Swarm! players, and Eve's active engagement with the trails further augments the map to highlight the relevance of those attractors. These attractors can in turn coordinate others to further document and engage an area, filling out the digital profile of regions that are of central use in human social behaviors, and effectively turning Swarm! players into an engineered team of self-directed, self-organized content curators. Every Swarm! player's behavior is thus influenced both by the structure of the game map, and the structure of the game map is influenced by the behavior of Swarm! players. However, since the initial structure of the Swarm! game map is dictated by the antecedent behavior of Swarm! players, this mechanic only serves to reinforce extant patterns of behavior.

---

[7] An *attractor* is just a location or state in a system toward which nearby states or locations tend to be "sucked." Minimum-energy states in mechanical system are commonly attractors. For instance, in a system consisting of a marble confined to the inside of a mixing bowl, the state in which the marble is at rest at the bottom of the bowl is an attractor: no matter where you start the marble, it will eventually end up at rest at the bottom of the bowl. For an accessible introduction to the language of attractors and dynamical systems theory, see Strogatz (2001) and Morrison (2008).

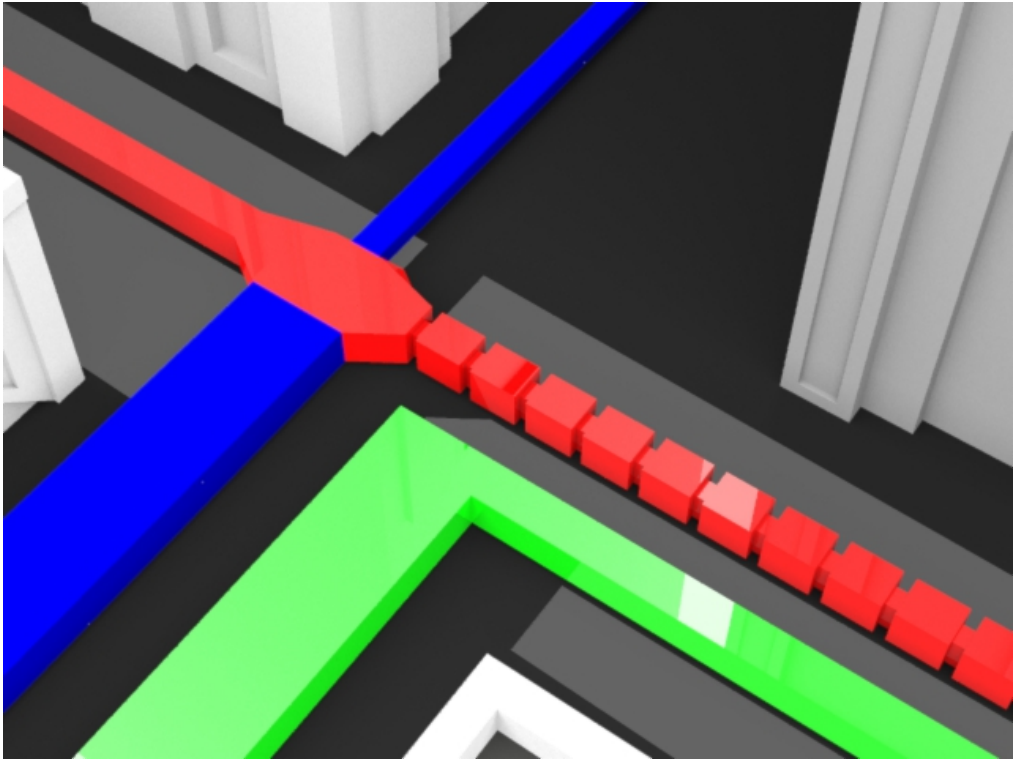

**Fig. 2** Our player Eve (indicated with the green trail) considers a regular interaction at a busy intersection with a hostile colony (red trail), which imposes caste-specific effects on a region.
Image credit: Kyle Broom

The resulting model highlights patterns of natural human behavior that can be directly harvested for computational work. For instance, consider the problem of locating a working electrical outlet at the airport[8]. Traditional resource distribution structures (like the financial markets or public regulatory structures) have until now failed to provide enough incentive to curate a digital outlet location map for wide public use, despite its potential value to both customers (who may wish to charge their electronics while they wait for a connecting flight), and the airport businesses (who might be able to draw customers and control the flow of airport patrons by advertising their location). Online databases like Yelp work well for services that have existing advocates, like restaurant owners, who can represent those interests by responding and reacting to Yelp reviews, but little incentive exists for a curation task like this. On the other hand, with suitable resolution Swarm! provides an immediate visual representation of the activity of airport patrons that allows for intuitive predictions about where the outlets might be: look for clustering behavior near walls. Moreover, Swarm! rewards active players for tagging public spaces with pictures and notes that fill in details of the interaction history at that location. The result is an NHC method for computing a solution to the problem of finding electrical outlets without the need for natural advocates or market representation to explicitly engineer this behavior.

---

[8] Credit goes to Robert Scoble for raising the example during a recent conversation about Swarm!.

This example has Swarm! players uncovering the use-value of objects which have been under-represented by other records of social value, and it has accomplished this without creating any additional demand on social behaviors. Perhaps a close analog is the use of GPS requests for identifying traffic congestion (Taylor, 2000), but the game mechanics of Swarm! generalizes the approach for a broad range of human activities. We turn now to a general discussion of the strategies described above.

## 2.2 NHC Applications of Swarm!?

Consider the mechanic described in Section **2.1** for modifying the game map by taking and tagging pictures. A strategically-minded Swarm! player will not use this ability at just any location (Rashid, 2006; Ames & Naaman, 2007); rather, she will study the structure of local trails over the course of a few days, and engage with the map in a tactically-optimal location--that is, a location that already experiences heavy traffic of the right sort. In this way, the Swarm! map will become a fairly detailed representation of patterns of player engagement with the real world; locations that are naturally highly trafficked will become increasingly important, and thus increasingly saturated with trails and in-game structures.

The fact that interesting locations in the game tends to mirror the interesting locations in the real world is central to Swarm!'s design. While Swarm!'s mechanics might well have some influence on the behavior of more strategically-minded players, that influence will *remain* a mirror of the aggregate pre-game behavior of the community, and thus a useful starting point for NHC data collection about use behavior. Ingress, a somewhat similar augmented reality game developed by Niantic Labs for Android mobile devices (Hodson, 2012), makes for an instructive contrast case. Ingress features two in-game "teams" (Enlightened and Resistance) involved in attempts to capture and maintain control of "portals," which have been seeded by Google at various real-world locations. Players take control of a portal by visiting the location (sometimes in cooperation with other players), and remaining there for a set amount of time. Players may also "attack" portals controlled by the opposing team through a similar location-based mechanic.

Notice the difference between tracking the behavior of Ingress players and tracking the behavior of Swarm! players. Despite both games featuring similar location-based mechanics, the fact that Ingress' portals--the significant in-game attention attractors--have been seeded by the game's designers renders the activity of Ingress players a poor proxy for their natural, out of game behavior, and thus a poor proxy for NHC data collection. In contrast, Swarm! players create the structure of the map themselves, and the strategically optimal approach to modifying it involves reinforcing existing patterns of behavior. The structure of the Swarm! map reveals at a glance sophisticated facts about the natural attention patterns of Swarm! players. It is this fact that makes Swarm! an important first step toward a mature NHC application.

Transitioning Swarm! from a NHC-oriented game to a real NHC application will involve tightly integrating Swarm!'s mechanics with real-world tasks. We suggest that Swarm!'s existing mechanics might be easily tied in to a service like Craigslist.org. Craigslist is a popular and free web-based service facilitating the exchange of good and services that run the gamut from used cars and furniture to prospective romantic encounters--all of which are organized geographically and easily searchable.  The Swarm! platform, with its built-in mechanics for tracking location, activity, and experience could serve as a platform for visualizing Craigslist service requests and evaluating the results of the transaction. If successful, such a system would allow for a self-organized, entirely horizontal resource and labor management system for its users. Such integration would be a large step toward turning Swarm! into the sort of robust economic HC application that we discuss in Section 4.

Consider the following hypothetical future in-game scenario: Eve, our intrepid player from Section 2.1, has access to a Craigslist-like service integrated with an advanced version of Swarm!, and this service informs her (on request) about posts made by other players in her immediate geographical region.  With access to this information, Eve can decide whether or not to accommodate the requests of other players in her vicinity.  Suppose, for instance, that Eve notices a posting near her home base requesting a 40 watt CFL light-bulb to replace a bulb that just burned out. Eve was targeted with the request because her patterns of behavior repeatedly cross paths with the requesting user; depending on how sophisticated the service has become, it might even recognize her surplus of light bulbs. In any case, Eve knows that she has several matching bulbs under her kitchen sink, and considers using the bulb to gain experience and influence within Swarm!. Eve notices that the specified drop point is on her way to work, and agrees to drop the bulb by as she walks to the subway. Perhaps the dropoff is coordinated by each party taking a picture of the object using QR codes that signal drop off and receipt of the object. Upon completion, this transaction augments player statistics within Swarm! to reflect the success of the transaction. As a result, Eve's public standing within the player community increases, just as it would have if Eve had participated in a coordinated attempt to seize a food source for her colony.  Her increased influence within game environment might increase the chances that her next request for a set of AA batteries is also filled.

This mechanic creates an environment in which contributing to the welfare of other Swarm! players through the redistribution of goods and services is rewarded not monetarily, but through the attraction of attention and the generation of influence and repute. The attention attracted by the request is converted into user experience upon completion of the task, allowing the user's behavior to have a more significant impact on the dynamics of the game. Again, this mechanic helps to blur the line between in-game and out-of-game interactions: the in-game world of Swarm! is a distillation and reflection of the everyday out-of-game world of Swarm!'s players.  Eve's history as a Swarm! player disposed to help other players in need might be intuitively presented to other members of her colony through features of her trail.  When Eve makes a request for aid other players will be more disposed to respond in kind.[9]

Although our examples have focused on minor transactions of relatively little significance,

---

[9] The influence of perceptions of fairness on economic interactions is an increasingly well-studied phenomenon among economists and psychologists.  For a comprehensive overview, see Kolm & Ythier (2006), especially Chapter 8  (Fehr & Schmidt).

the game mechanics described here suggest a number of important principles for designing HC systems that harvest the computational dynamics of natural human activity, and the profound impacts these applications might have on a number of vitally important human activities, including education, politics, and urban development. We focus the remaining discussion on economic considerations.

## 3. Naturally optimizing the economy

We can think of the global economy as being a certain kind of HC system in which the problem being computed involves the search for optimal (or near-optimal)[10] allocations of raw materials, labor, and other finite resources ("the economic optimization problem").  This approach to economic theory is broadly called "computational economics" (see e.g. Velupillai, 2000; Norman, 1996), and it takes economic theory to be an application of computability theory and game theory.  Historically, some economists have argued that a free capitalist market composed of minimally constrained individual agents (and suitable technological conditions supporting their behavior) provides the most efficient possible economic system (Hayek, 1948). We shall conclude our paper with a discussion of NHC applications as an alternative approach for tackling the economic optimization problem.

Kocherlakota (1998) argues that money is best thought of as a "primitive form of memory" (*ibid.* p. 2).  That is, money is a technological innovation that provides a medium for a limited recording of an agent's history of interactions with other agents.  On this view, rather than being an intrinsic store of value or an independent medium of exchange, money is merely a way to record a set of facts about the past.  Kocherlakota argues that this technological role can be subsumed under "memory" in a more general sense, and that while access to money provides opportunities for system behavior that wouldn't exist otherwise, other (more comprehensive) kinds of memory might do all that money does, and more: "...in at least some environments, memory [in the form of high quality information storage and access] may technologically dominate money" (*ibid.* p. 27).

If this characterization is correct, then solving the economic optimization problem involves accomplishing two distinct tasks: identifying precisely *what* information should be recorded in economic memory, and we must devising ways to store and manipulate that information.  We might understand Yelp as recording user accounts of a service that attempts to meet these memory challenges. Yelp users leave comments, reviews, and ratings that provide a far more detailed and relevant transaction history with customers than is represented by the relative wealth of the business as a market agent. Luca (2011) finds not only that these reviews have an impact on revenue, but that impact is strengthened with the information content of the reviews, suggesting one place where money may be showing evidence of domination by rich sources of memory.

---

[10] The definition of "optimal" is disputed, but the discussion here does not turn on the adoption of a particular interpretation.  In general, recall that solving the economic optimization problem involves deciding on a distribution of finite resources (labor, natural resources, &c.).  Precisely which distribution counts as "optimal" will depend on the prioritization of values.  A robust literature on dealing with conflicting (or even incommensurable) values exists.  See, for example, Anderson (1995), Chapter 13 of Raz (1988), and Sen (1997).

Swarm! offers a natural approach for meeting the same challenges, in which NHC is leveraged to help solve the economic optimization problem without introducing new economic frictions. This computational work is accomplished through the recording of trails that represents incremental changes in the use history of that location. As Swarm! maps become increasingly detailed and populated they likewise come to function as an effective representation of the attention economy (Simon, 1971; Weng, 2012) in which the saturation of trails around an object approximates a quantitative measure of the value of objects relative to their use[11]. We treat this measure as the aggregate "use-value" of the object (Vargo et al., 2008), and argue that a model of the use-value of objects allows for novel NHC-based solutions to a variety of standard problems in the optimization of economic systems. A full articulation of the attention economy is not possible here, but we will provide a sketch of one possible implementation using the Swarm! framework.

## 4. Developing the Attention Economy

Recall the central mechanic of Swarm!. GPS data about players' movement patterns are aggregated, whether or not a player is actively engaged with the game. Strategically-minded players are rewarded for tagging and modifying the map in a way that gives rise to a detailed reflection of how all Swarm! players use the space covered by the map. The data collected by a Swarm!-like application has the potential to encode many of the facts that might otherwise be encoded less explicitly. Monetary transaction records act as proxy recordings for what we have called *use-value*. The mechanics of Swarm! suggest a way to measure use-value directly by recording how economic agents move through space, how their movement is related to the movement of others, what objects they interact with, the length and circumstances of those interactions, and so on. By tracking this data, we can transform the everyday activities of agents into records of what those agents value and to what degree. This is the "high quality information storage and access" that Kocherlakota suggests may come to technologically dominate currency as economic memory. Still, a number of practical challenges must be surmounted before a NHC/AE based approach to solving the economic optimization problem is realistically viable.

Any implementation of an attention economy in which the economic optimization problem is solved with NHC will clearly involve data collection on a scale that goes far beyond what's possible in Swarm! or with Google Glass, as the mere tracking of gross geospatial position will not record enough information to (for instance) assay the value of individual material objects like pens and lightbulbs. Swarm! is an incremental step in that direction, with the more modest and technologically feasible goals of acclimating people to regular engagement with AE platforms, and with developing the social norms appropriate to the demands of an AE. The structure of human communities is strongly coupled to the technology available during their development. Absent major catastrophes, the sort of ubiquitous computing and social norms necessary for the implementation of an AE will continue to develop in tandem.

---

[11] As opposed to value relative to *exchange*. See Marx (1859).

Indeed, the success of AE in some sense depends on the development of social customs and attitudes to compensate for the more invasive social coordination technologies that dominated the Industrial Age, which are almost universally characterized by the establishment of hierarchical institutions of control. In such a system, power is concentrated in the hands of the very few, to be executed within very narrow channels of operation. For the disenfranchised, finding ways to circumvent or usurp this power is often a more attractive than accumulating power through so-called "legitimate" means--especially as the powerful increasingly protect their positions through deliberate corruption and abuse, thereby weighting the system heavily against "fair play". In other words, enterprising opportunists looking for success in systems with limited hierarchical control have a disproportionate incentive to "game the system", or exploit loopholes in the rules in ways that give them a disproportionate advantage. Preventing the exploitation of such loopholes requires an ever increasing concentration of power, creating greater incentives to break the system, and greater costs for failing in those attempts. Social customs discouraging such behavior must be imposed from the top, often with violence, as a means of retaining control, since these customs are not reinforced from below.

In contrast, the AE describes a self-organizing system without hierarchical control or concentrations of power, because the rules for operating within the system also support the success of the system as a whole, and so are supported from the bottom without need for top-down enforcement. In other words, the impulse to game an attention economy can be actively encouraged by all parties, since individual attempts to gain a disproportionate advantage within the system simultaneously reinforce the success of the system overall. Recall from section 2.1, when Eve snaps a picture of a highly trafficked block. This apparently self-interested act to improve her own in-game resource collection rate is simultaneously a contribution to the economic optimization problem, and is therefore reinforced by her colony's goals. Of course, Eve is not only rewarded by pursuing self-interested goals; potentially everything Eve does in an attention economy is computationally significant for her community, and therefore her community can support Eve in the pursuit of any goals she wishes without worrying about how her actions might upset the delicate balance of power that supports institutional control. In an attention economy, Eve is not rewarded to the extent that she appeals to existing centers of power; instead, she is rewarded to the extent that her participation has an impact on the development of her community.

We conclude by mentioning some design considerations inspired by Swarm! for building an "Internet of Things" that facilitates the the use of NHCs for managing the attention economy. Most obviously, Swarm! is a step toward the creation of pervasive, universally accessible, comprehensive record of the relationship between agents, locations, and objects. As we have said, widespread location and identity tracking of at least *some* sort is essential for the implementation of a true AE.  This is a major design challenge in at least two senses: it is a technical engineering challenge, and a social engineering challenge.

The solution to the first challenge will still require technological progress; we do not yet have ubiquitous distribution of the sort of computing devices that would be necessary to implement the fine-grained level of data collection that a real AE would require. In addition to aggregate movement patterns, an AE platform will need to track patterns in the relationships between agents and physical objects.  Sterling (2005) introduces the term "spime" to refer to inanimate objects that are trackable in space and time, and broadcast this data throughout their lifetimes. Objects that are designed to live in an attention economy must track more than just their own location and history: they must be able to track their own use conditions, and change state when those use conditions have been met.  This will require objects to be sensitive not just to their own internal states, but also to the states of the objects (and agents) around them: this is the so-called "Internet of Things" (Atzori et al., 2010). There is already some precedent for very primitive functions of this sort.  Consider, for instance, the fact that modern high-end televisions often feature embedded optical sensors to detect ambient light levels, and adjust backlighting accordingly for optimal picture quality.  We can imagine expanding and improving on that kind of functionality to develop (say) a television that mutes itself when the telephone rings, pauses when you leave the room, or turns itself off when a user engages deeply with another object (for instance a laptop computer) that's also in the room. These examples are relatively mundane, but they are suggestive of the sort of industrial design creativity and integration needed to design AE-optimized artifacts.

Swarm! approaches this design challenge by imposing some novel clustering methods represented by the caste and colony system.  The colony system is a geographical constraint designed to cluster colony members to ensure that they aren't spread so thin as to undermine the game dynamics. The caste system is a design constraint on the patterns of user activity, and allows users to tell at a glance the functional results of some possible sequence of engagements without knowing too many details about other players. This latter feature is inspired directly by ant colonies, and is important to the organizational dynamics of an AE. In particular, it gives contexts in which it is appropriate for certain agents to have disproportionate influence  on some computing task, thereby carving out emergent hierarchies and cliques. The AE/NHC platform is thus applicable to the solution of non-economic social problems, and can be leveraged to help compute solutions to other legal, political, and social puzzles.

As an illustration of how NHCs might be applied to the distribution and management of resources and labor, consider the transaction history for some arbitrary object X. If this record has been reliably maintained on a user-per user basis, it might serve as the basis for resolving disputes about ownership, rights of use, and other coordination problems traditionally settled by legal and political frameworks. If I have years of history driving a specific car on Wednesday mornings, and the use record shows you driving this car some particular Wednesday morning, then absent some explanation this appears to be a disturbance in use patterns. This information might itself be enough to warrant a complaint through official channels and initiate the machinery of the public justice system to account for this disturbance. In other words, a well-maintained record of the use history of an object might serve as a foundation for NHC solutions to political and legal disputes, and provides a framework for dealing naturally with apparent cases of "stealing" without requiring anything like the disruptive technologies of property, contracts, and other legal frictions.

This is the real heart of the AE/NHC approach to economic optimization: the NHC acts entirely upon data about local patterns of attention, use, and interaction without significantly disturbing the behavioral patterns that generate the data. Rather than indirectly recording facts about my contribution to (or value of) some object or process in monetary memory, which requires its own set of social conventions and techniques to maintain, those facts are recorded *directly* in the history of my relationship to the object or process. We suggest that careful management of those facts, combined with a distributed NHC framework, might allow for a far more efficient economic system than any money-based system.

We've given a characterization of the shape and character of the first of the two design challenges we mentioned above: the technical engineering challenge. While solving this challenge is central to the implementation of the AE, we should not overlook the importance of solving the second challenge either. While technological advances are important, so are advances in the relationship between humans, technology, and society at large. Just as the dissemination of other major, epoch-defining technologies (like the automobile or the telephone) were accompanied by a certain degree of widespread anxiety and social disruption, we expect that the adoption of the ubiquitous computing platforms required for AE implementation (and their concomitant changes in social practice) will be associated with some unrest as society acclimates to some of the underlying changes. In this respect, Swarm! is more than just an experiment in designing a NHC application--it is an attempt to give society at large a chance to experience the artifacts and socio-cultural practices required for a well-managed AE. The more time we have to grapple with those issues as a community, the smoother the transition to the future will be.